\documentclass{iopart}
\usepackage{graphicx}
\usepackage{bm}

\usepackage{amsbsy}
\usepackage{amssymb}
\usepackage{iopams}  

\def\dd{\mbox{d}}

\def\S{{\bf S}}

\begin{document}


\title[Pulling Membrane Tethers]{Models of dynamic extraction of lipid tethers from cell membranes}

\author{Sarah A. Nowak}
\address{RAND Corporation, Los Angeles, CA 90405}

\author{Tom Chou}
\address{Dept. of Biomathematics, UCLA, Los Angeles, CA 90095-1766}
\address{Dept. of Mathematics, UCLA, Los Angeles, CA 90095-1555}

\date{\today}


\begin{abstract}


When a ligand that is bound to an integral membrane receptor is
pulled, the membrane and the underlying cytoskeleton can deform before
either the membrane delaminates from the cytoskeleton or the ligand
detaches from the receptor. If the membrane delaminates from the
cytoskeleton, it may be further extruded and form a membrane tether.
We develop a phenomenological model for this processes by assuming
that deformations obey Hooke's law up to a critical force at which the
cell membrane locally detaches from the cytoskeleton and a membrane
tether forms.  We compute the probability of tether formation and show
that they can be extruded only within an intermediate range of force
loading rates and pulling velocities.  The mean tether length that
arises at the moment of ligand detachment is computed as are the force
loading rates and pulling velocities that yield the longest tethers.
%
%
\end{abstract}


\ead{tomchou@ucla.edu}

\section{INTRODUCTION}
\setcounter{footnote}{0}

Adhesion between cells plays an important role in a number of
biological processes involving cell motility and cell-cell
communication.  Cell-cell adhesion is mediated by integral membrane
proteins on the surfaces of interacting cells.  Cadherins, which
mediate binding between cells of the same type within a tissue, bind
to themselves, while integrins, which mediate binding between
different cell types bind to Inter-Cellular Adhesion Molecules (ICAMs)
or Vascular Cell Adhesion Molecules (VCAMs) \cite{Lodish-03}.
Understanding the physics of this adhesive interaction requires an
understanding of both the protein-protein bond as well as the cell's
mechanical response when these bonds become stressed.  If forces act to
pull the two cells apart, either of the cells' cytoskeletons and
plasma membranes may deform. Under certain conditions, the lipid
membranes can delaminate from the underlying cytoskeleton and be
pulled into long tethers. At any time during this process, the bonds
holding the two cells together may also break, arresting tether
extraction.

A specific biological process in which bond dissociation and membrane
deformation must be considered simultaneously is leukocyte
extravasation, which is part of the process by which leukocytes are
recruited to inflamed or infected tissue. Endothelial cells that make
up blood vessels preferentially express cellular adhesion molecules,
including selectins, near wounded tissue. Leukocytes circulating in
the blood can then bind to the endothelial cells via their own cell
surface proteins. Bonds between the leukocytes and the endothelial
tissue are transiently made and broken as a shear flow in the blood
vessel pushes the leukocyte, rolling it across the endothelial layer
\cite{UDO}.  The rolling leukocytes contain microvilli that are
enriched in adhesion molecules that preferentially attach to
endothelial cells. During rolling, these microvilli tethers can extend
under the hydrodynamic shear force of blood flow in the vessel
\cite{ShaoHochmuth98}. At the same time, forces imposed on the
endothelial membrane via the adhesion molecule can cause the
endothelial membrane to form a tether \cite{GirdharShao07}. Tether
formation and extension of microvilli can decrease the force that the
adhesive bond feels from the shear blood flow.

Both the physics of cell membrane deformation and the mechanics of
ligand-receptor bonds have been studied extensively.  Micropipette,
atomic force microscope (AFM), optical trap, and magnetic bead
techniques have been used to pull membrane tethers from cells and
probe the properties of the cell membrane \cite{SCH09}.  The diameter
of the extracted tether depends on the membrane surface tension and
bending rigidity and, if the tether is being extended at a constant
velocity, the membrane viscosity \cite{HochmuthEvans82, JDai03011995}.
Therefore, these quantities can be inferred from tether pulling or
poking experiments \cite{FYGENSON}. Theoretically, Euler-Lagrange
methods have been used to compute equilibrium tether shapes and the
force-extension curve for a tether pulled quasi-statically from a
lipid vesicle \cite{Powers-02,DER02}. These theoretical models of pure
lipid bilayers show only a $\sim 10$\% overshoot, or barrier, in the
force-extension curve before a tether is extracted from an
asymptotically flat membrane \cite{Powers-02,DER02}\footnotemark, some
experiments show a significant force barrier to tether formation
\cite{SunGrandbois05,KosterDogterom05,NASSOY}.  These large force
barriers to membrane tether formation arise in living cells and is
attributed to membrane adhesion to the underlying actin cytoskeleton
\cite{SunGrandbois05,BenoitGaub00}.  When tethers are pulled from
giant artificial vesicles, the size of the force barrier can increase
only when the area on which the pulling force is exerted is increased
\cite{KosterDogterom05}. For smaller vesicles, area and volume
constraints may also influence the tether force-extension relationship
\cite{ANA}.  These may arise from nonlocal terms in the functional
describing the lipid membrane energetics. For example, area-difference
elasticity can give rise to a restoring forces that continually
increase as tether length increases \cite{GLA06,RAP96}.  Such nonlocal
effects will be important only when the tether comprises an appreciable
fraction of the total membrane area.  Force curves that do not
saturate at long tether extensions can also arise when part of the
membrane reservoir adheres to a substrate \cite{NASSOY}. The relative
importance of nonequilibrium forces arising from the viscosity of both
the membrane lipids and surrounding solution have also been estimated
\cite{Hochmuth-96}.

\footnotetext{The barrier represents the energy that must be
overcome before a lipid tether can be drawn out. The $\sim 10\%$
energy barrier arises from the Helfrich free-energy-minimizing
geometry of an incipient lipid membrane tether in the absence any
cytoskeletal attachments\cite{Powers-02,DER02} (cf. dashed curve in 
Fig. \ref{schematic}}

During tether pulling experiments, and in their corresponding
theoretical models, the molecular bond attaching the pulling device to
the membrane is assumed to always remain intact. However, a typical
ligand-receptor bond used to connect the pulling device to the lipid
membrane (and possibly the underlying cytoskeleton) can rupture upon
pulling. Although the details of a bond's energy landscape can be
probed using dynamic force spectroscopy
\cite{MerkelEvans99,Evans01,HeymannGrubmuller00, DudkoUrbakh03}, one
can usually assume that bond rupturing is dominated by a single
activation barrier that can be lowered by an externally applied
pulling force. Most AFM studies of molecular strength are performed on
proteins that have been isolated from cells.  However, recent studies
have probed bond strengths between proteins still embedded in a live
cell membrane \cite{HanleyKonstantopoulos03,SchmitzGottschalk08}.
While using live cells has the advantage that the post-translational
modifications of membrane proteins are preserved, the mechanical
deformation of the cell's cytoskeleton and membrane must also be taken
into account. To model this system, a viscoelastic Kelvin model was
used to fit experimental measurements of the force-extension
relationship to determine effective cellular adhesion
\cite{SchmitzGottschalk08}.


In this paper, we develop a dynamic model that incorporates
phenomenological theories of membrane and cytoskeleton deformation,
tether extraction, and the kinetics of ligand-receptor detachment. In
contrast to an equilibrium model determining tether extraction and
detachment from an adhered vesicle \cite{ANA}, we find relationships
that define when tether extraction is likely, and the typical length
of the tether pulled before the ligand-receptor bond ruptures. In the
next section, we motivate a simple mechanical model using
phenomenological forms for the force-extension relationship of the
membrane.  Given the large bending energies of lipid bilayer
membranes, and the relatively strong attachment of membranes to
cytoskeleton, we will neglect thermal fluctuations of the membrane,
but implicitly include thermally-driven ligand-receptor bond
dissociation. Dynamical equations are written for two commonly
employed experimental protocols, a linear force loading rate, and a
constant bond pulling speed.  In the Results and Discussion, we
compute the probability of tether formation and plot universal curves
that delineate regimes where tethers are likely to form.  Mean tether
lengths are also plotted for both pulling protocols.

\section{Mathematical Model}

\begin{figure}[h!]
\begin{center}
\includegraphics[width=2.9in]{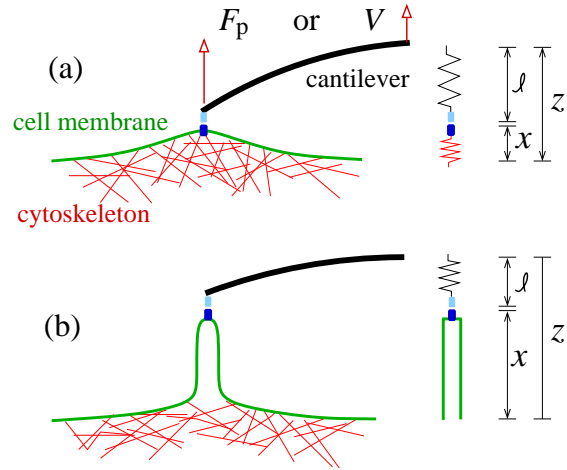}
\end{center} 
\vspace{-2mm}
\caption{(a) A cell membrane can be pulled by a device such as a
cantilever.  The device can be moved at fixed velocity $V$, or, at a
specified force $F_{\rm p}(t)$. Under specified force conditions, the
transduction of force through the device is assumed instantaneous such
that $F_{\rm p}(t)$ is always acting across the ligand-receptor bond.
Before membrane delamination from the underlying cytoskeleton, the
pulling results in small membrane and cytoskeletal deformations. (b)
After membrane delamination, the pulling device can extrude long lipid
tethers without deforming the cytoskeleton.  The total system length
$z$ is the sum of the displacement $x$ of the membrane protein from
its initial position and $\ell$, the increase in the length of the
cantilever spring from its unstretched length.}
\label{FIG1} 
\end{figure}

Consider the system depicted in Figure \ref{FIG1}.  A ligand is bound to
an integral membrane protein, which may also be directly associated
with the cytoskeleton.  The ligand may be  attached to a pulling device via
a cantilever spring and is pulled with either a fixed speed, or a force that
increases linearly in time.

As the ligand is pulled, the cytoskeleton first deforms, and
eventually can detach from the membrane.  At this point, the lipid
membrane may flow into a tether.  At any point during the
cytoskeletal or membrane deformation, before or after
membrane-cytoskeleton delamination and tether formation, the ligand
affixed to the pulling device may detach from the membrane receptor
protein.



\subsection{Membrane Mechanics}

We first consider the response of the membrane-cytoskeleton system to
an externally applied pulling force $F_{\rm p}(t)$.  The rate at which
the receptor-ligand complex moves will be described by

\begin{equation}
\frac{\dd x(t)}{\dd t} = -\xi\left[\frac{\partial
U(x)}{\partial x}\bigg|_{x=x(t)}-F_{\rm p}(t)\right]
\label{eq:xDot}
\end{equation}
where $\xi$ is a mobility that is inversely related to the viscosity of the
membrane lipid \cite{Hochmuth-96}. In general, although $\xi$ depends
on the configuration of the system defined predominantly by $x(t)$, we
neglect the details of this dependence and assume it to be  constant. 

The term $U(x)$ represents the energetic cost associated with
deforming the cell membrane and underlying cytoskeleton when the
receptor is displaced a distance $x$ normal to the flat membrane.
This phenomenological energy can be derived from a detailed
consideration of the membrane-cytoskeletal mechanics. For simplicity,
we will assume the membrane mechanics are governed by a Helfrich
free-energy \cite{MIA94} that includes a lipid bilayer bending rigidity
and an effective thermally-derived entropic membrane tension.
We will assume that the membrane reservoir is large enough such that 
an extruded tether negligibly depletes the reservoir. 
Hence, global contributions to the membrane energetics,
such as area-difference elasticity, can be neglected.  

Experiments in which tethers were pulled from live cells found a
significant force barrier to tether formation \cite{SunGrandbois05,
BenoitGaub00}.  While a smaller force barrier can also arise in pure
lipid membranes \cite{Powers-02}, we will assume that the plasma
membrane is attached to an underlying cytoskeleton (with anchoring
molecules), which we model as a linear elastic material provided the
deformation is small. The receptor that binds the ligand that is
attached to the pulling device can also be a transmembrane receptor
that is directly attached to the cytoskeleton.  As the membrane is
initially pulled, the cytoskeleton will elastically deform as a
Hookean spring.  The receptor or anchoring molecules will break at a
deformation $x_{0}$, and the lipid membrane will be drawn into a
tether. This occurs at a critical delamination force $F_{\rm c}$.
Thus, for displacements $x<x_{0}$, where the Hookean approximation for
the membrane-cytoskeleton assembly is valid, the membrane-cytoskeleton
carries an effective spring constant $F_{\rm{c}}/x_0$, where $F_{\rm
c}$ is the critical delamination force.  Experimentally, the
cytoskeleton typically detaches from the membrane where the
filament-free tether forms \cite{AFR06,BorghiWyart07}; therefore we
can assume a linear force extension relationship of the form

\begin{equation}
{\partial U\over \partial x} = {x\over x_0}
F_{\rm{c}}, \quad x<x_{0}.
\end{equation}
At a displacement $x=x_{0}$, the membrane delaminates from the
cytoskeleton and a lipid membrane tether forms. Under the infinite
reservoir and simple Helfrich free-energy assumption, the tether can
then elongate indefinitely under a constant force $\partial U/\partial
x = F_0$ that is intrinsic to the lipid tether and is determined by
the membrane bending rigidity $\kappa$ and entropic surface tension
$\sigma$ \cite{Powers-02,DER02}: $F_{0} = 2\pi\sqrt{2\kappa\sigma}$.
\begin{figure}[h!]
\begin{center}
\includegraphics[width=2.9in]{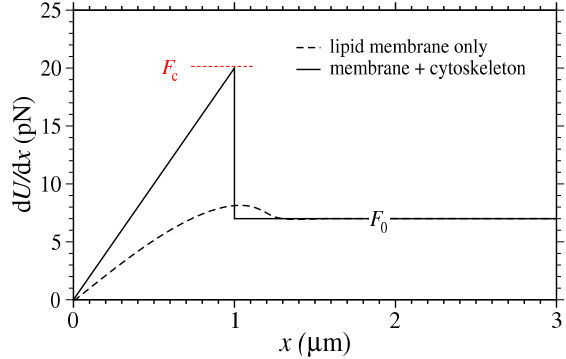}
\end{center} 
\vspace{-2mm}
\caption{We assume that when the receptor is initially associated with
the cytoskeleton, $\partial_{x}U(x)$ increases linearly until some
maximum force $F_{\rm{c}}$ in the barrier to tether formation is
reached, after which the tether extends with constant force $F_0$
(solid curve).  The qualitative features of this force-extension curve
is observed in numerous systems \cite{JDai03011995,
SunGrandbois05,KosterDogterom05,NASSOY,BenoitGaub00}. For reference,
we show $\partial_{x}U(x)$ as a function of $x$ when the only
contributions to $U(x)$ come from membrane surface tension and bending
rigidity (dashed curve).  This curve was calculated using the method
described in \cite{Powers-02} assuming a membrane bending rigidity of
20$k_{B}T$ and a surface tension of 0.0076 dynes/cm.}
\label{schematic} 
\end{figure}
The phenomenological membrane force-displacement relationship for both
attached (solid curve) and free (dashed curve) membranes are shown in
Figure \ref{schematic}.  The barrier $F_{\rm c}$ to tether formation
is larger for receptors or membranes that are attached to the
underlying cytoskeleton, than for the free lipid membrane case.

In order to close the equations of our basic model, we must specify
$F_{\rm p}(t)$.  Henceforth, we will consider two cases typically
realized in experiments: a linearly increasing (in time) pulling
force, and a fixed pulling speed.

\subsection{Linear Force Ramp}

For a force linearly increasing in time, $F_{\rm p}(t) = \Gamma
t$, where $\Gamma$ is the rate with which the 
force increases.  Eq. \ref{eq:xDot}
becomes
\begin{equation}
{\dd x(t) \over \dd t} = -\xi\left[{\partial U(x) \over \partial x} -
\Gamma t\right].
\label{eq:XDOTRAMP}
\end{equation}
Upon defining the time $t_{0}$ at which the membrane delaminates from
the cytoskeleton provided the ligand-receptor bond has not yet
ruptured by $x(t=t_{0}) \equiv x_{0}$, we have ${\partial U\over
\partial x} = (x/x_0)F_{\rm{c}}$ for $t\leq t_{0}$, and ${\partial
U\over \partial x} = F_{0}$ for $t>t_{0}$. Thus, Eq. \ref{eq:XDOTRAMP}
is solved by

\begin{equation}
x(t<t_{0}) = {x_{0} \Gamma \over F_{\rm c}}t - 
{\Gamma x_{0}^{2} \over \xi F_{\rm c}^{2}}(1-e^{-\xi F_{\rm c}t/x_{0}}),
\end{equation}
and
\begin{equation}
\begin{array}{ll}
x(t>t_{0}) = & x_{0} - \xi F_{0}(t-t_{0}) \\[13pt]
\: & \hspace{3mm} \displaystyle + {\xi \Gamma \over 2}(t^{2}-t_{0}^{2}), 
\end{array}
\end{equation}
where the delamination $t_{0}$ is determined from the solution to

\begin{equation}
{\Gamma \over F_{\rm c}}t_{0} - {\Gamma x_{0} \over \xi 
F_{\rm c}^{2}}(1-e^{-\xi F_{\rm c}t_{0}/x_{0}}) = 1.
\label{T0RAMP}
\end{equation}

\subsection{Constant Pulling Speed}

In the case of constant pulling speed, we must include the dynamics of
the device deformation $\ell(t)$.  Since the velocity of the pulling
device is fixed, we note that the total displacement $z(t) = x(t) +
\ell(t)$ obeys

\begin{equation}
{\dd z\over \dd t} = \frac{\dd x}{\dd t}+\frac{\dd \ell}{\dd t} = V.
\label{eq:pullingC}
\end{equation}
This equation holds only when the ligand is attached to the
membrane-bound receptor.  Let us assume that the pulling device has an
internal response that is modeled by a simple spring so that the force
$F_{\rm p}(t)$ that the spring exerts on the ligand is

\begin{equation}
F_{\rm p}(t) = K \ell(t), 
\label{eq:pullingForce}
\end{equation}
where $K$ is the spring constant of the pulling device.  Since the
pulling force is proportional to $\ell(t)$, it will ultimately depend
on the pulling rate $V$ and the physical properties of the pulling
device (represented by an elastic cantilever in Figure 1) and cell
membrane through Eq. \ref{eq:pullingC}.  Upon integrating Eq.
\ref{eq:pullingC} and using the initial conditions
$x(t=0)=\ell(t=0)=0$, we find

\begin{equation}
x(t) = Vt-\ell(t).
\label{XL}
\end{equation}
Substituting  Eqs. \ref{XL} and  \ref{eq:pullingForce} into Eq. \ref{eq:xDot}, and
expressing the dynamics in terms of the device deformation, we find a
closed equation for $\ell(t)$:

\begin{equation}
\frac{\dd \ell}{\dd t} =V + 
\xi\left[\frac{\partial U(x)}{\partial x}\bigg|_{x=Vt-\ell(t)}-K\ell(t)\right].
\label{lDot}
\end{equation}
This equation is solved by 

\begin{equation}
\begin{array}{l}
\displaystyle \ell(t\leq t_{0}) = {VK x_{0}^{2} \over \xi (F_{\rm
c}+ Kx_{0})^{2}}(1-e^{-\xi(K+F_{\rm c}/x_{0})t}) \\[13pt]
\:\hspace{4.3cm} \displaystyle + {F_{\rm c} Vt \over F_{\rm
c}+Kx_{0}}
\end{array}
\label{eq:LofTsmallT}
\end{equation}
for $t<t_0$, and 
\begin{equation}
\begin{array}{ll}
\ell(t>t_{0}) & \displaystyle = {V+\xi F_{0}\over \xi K}(1-e^{-\xi K (t-t_{0})}) \\[13pt]
\: & \hspace{1cm}+ \ell(t_{0})e^{-\xi K (t-t_{0})}
\end{array}
\label{eq:LofTbigT}
\end{equation}
for $t>t_{0}$ (when $\partial U/\partial x = F_{0}$).  Here, the time
$t_0$ at which tether formation occurs is found by evaluating
Eq. \ref{XL} at time $t=t_{0}$, $x(t_{0})=x_{0}=Vt_{0}-\ell(t_{0})$,
yielding an implicit equation for $t_{0}$:

\begin{equation}
\begin{array}{l}
\displaystyle V t_0- x_0 = {F_{\rm c} Vt_{0} \over F_{\rm
c}+Kx_{0}} \\[13pt] \:\hspace{1.5cm} \displaystyle + {VKx_{0}^{2}
\over \xi (F_{\rm c}+ Kx_{0})^{2}}(1-e^{-\xi(K+F_{\rm
c}/x_{0})t_{0}}).
\label{T0}
\end{array}
\end{equation}
After evaluating $t_{0}$ numerically, $\ell(t)$ is found in terms of
the $K$, $V$, $\xi$, $F_{\rm c}$, $F_{0}$, and $x_{0}$, and 
the membrane displacement can be found using Eq. \ref{XL}.

\subsection{Ligand-Receptor Dissociation}

The dynamics described above for the membrane and pulling device
deformations assume that the pulling device remains attached to the
membrane through an unbroken ligand-receptor bond.  Since all external
forces are transduced through the ligand-receptor bond, the pulling
force $F_{\rm p}(t)$ on the membrane (cf. Eq. \ref{eq:xDot})
vanishes once the ligand-receptor bond ruptures.  However, the probability of
ligand-receptor bond dissociation itself depends on the applied force
$F_{\rm p}(t)$.  We can model the breaking of the ligand-receptor bond by
a Poisson process and define a ligand-receptor bond survival probability
$Q(t)$ that obeys 
\begin{equation}
\frac{\dd Q(t)}{\dd t} = -k_{\rm r}(t)Q(t),
\label{eq:dQdT}
\end{equation}  
where $k_{\rm r}(t)$ is the force-dependent rupture (or dissociation)
rate of the ligand from the receptor. We assume that $k_{\rm r}(t)$
takes a simple Arrenhius form \cite{BELL0}:

\begin{equation}
k_{\rm r}(t) = k_0 e^{F_{\rm p}(t)d/k_{B}T},
\label{RATE}
\end{equation}
where $d$ is the length of the ligand-receptor bond and $k_{B}T$ is
the thermal energy.  The solution to Eq. \ref{eq:dQdT} is explicitly

\begin{equation}
Q(t) 
= \exp\left[-k_{0}\int_{0}^{t}
e^{F_{\rm p}(t')d/k_{B}T}\dd t'\right].
\label{QT}
\end{equation}
More complex models of dynamics bond rupturing can be derived
\cite{EVANS,BELL}.  Here, for simplicity, molecular details such as the
thermally-induced bond-breaking attempt frequency and the intrinsic
free energy of the unstressed ligand-receptor bond are subsumed in the
effective rate parameter $k_{0}$.

Since $w(t) \equiv -\dd Q(t)/\dd t = k_{\rm r}(t)Q(t)$ is the bond
rupture time distribution,
%
the mean membrane displacement at the time of ligand-receptor rupture
(the mean maximum displacement) is given by

\begin{equation}
\langle x^{*}\rangle \equiv  \int_{0}^{\infty}w(t)x(t)\dd t =  \int_{0}^{\infty}k_{\rm r}(t)Q(t)x(t)\dd t.
\label{meanX}
\end{equation}

In our subsequent analysis, we will combine bond rupturing statistics
with membrane tether dynamics and explore, as a function of the
physical parameters, the probability of tether formation, and the
length of pulled tethers should they form.  To be concrete, we will
use typical parameter values (listed in Table I) found from the
relevant literature to guide our analysis.

\vspace{3mm}

\begin{tabular}{|c|c|l|}
\hline parameter&range of values & reference\\ \hline $d$ & 0.8-1.0 nm
&\cite{Leckband95} (streptaviden-HABA)\\ \hline $\xi$ & $\sim 1
\mu$m/(pNs) &\cite{Hochmuth-96}\\ \hline $x_0$ & $\sim 1-4
\mu$m&\cite{SunGrandbois05}\\ \hline $F_0$ & 3-380pN &
calculated\footnotemark
\\ \hline \cline{1-2}{$F_{\rm c}$} & 
100-380pN & \cite{BorghiWyart07} (red blood cells) \\ &
1-100pN &\cite{SakoKusumi98} (epithelial cells) \\ \hline 
$K$ & 8-11pN/$\mu$m &
\cite{SunGrandbois05}\\ \hline $k_0$ & 10$^{-5}$-10/s &
\cite{WardHammer95}\\ \hline  $V$ & $\sim 3$
$\mu$m/s & \cite{SunGrandbois05}\\ \hline
$\Gamma$ & 10pN/s & \cite{SunGrandbois05} \\ \hline  
\end{tabular}

\footnotetext{$F_0 = 2\sqrt{2}\pi\sqrt{\kappa\sigma}$ where $\kappa$
is the membrane bending rigidity and $\sigma$ is the effective
membrane surface tension\cite{Powers-02}.  We assumed $\sigma =
3-1200$pN/$\mu$m\cite{MorrisHomann01} and $\kappa = 10-20k_{B}T$
\cite{Evans-90}}

%
%

\section{Results}

Here, we compute the dynamics of tether formation and ligand-receptor
bond rupturing under both linear force loading and constant pulling
velocity protocols.

\subsection{Linear Force Ramp}

\begin{figure}[h!]
\begin{center}
\includegraphics[width=4.0in]{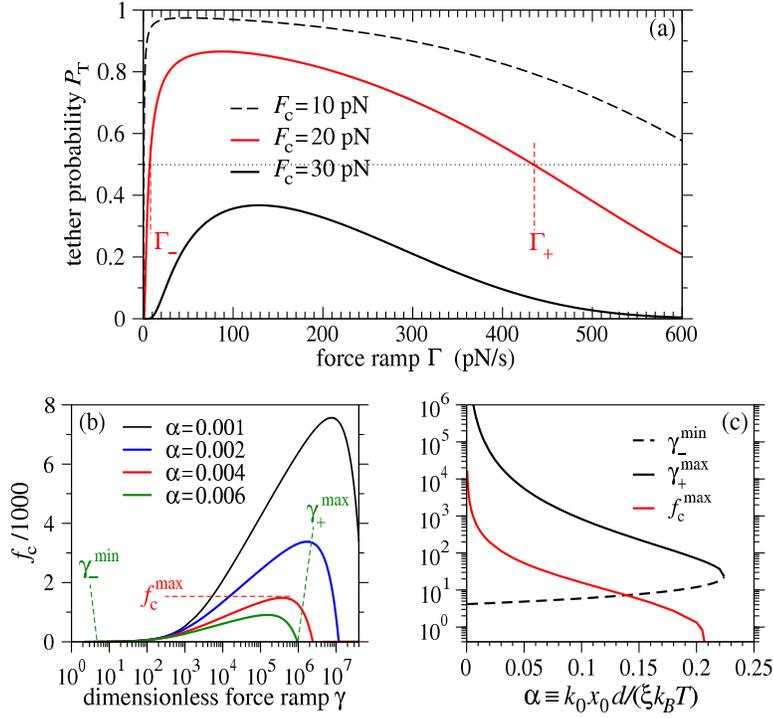}
\end{center} 
\vspace{-2mm}
\caption{(a) Tether formation probability $P_{\rm T}\equiv Q(t_{0})$
as a function of force ramp rate $\Gamma$. Note that results for the
constant force ramp protocol are independent of the free tether
restoring force $F_{0}$. $\Gamma_{\pm}$ denote the linear loading
rates at which the tether formation probability $P_{\rm T} = 1/2$.
Applied loading rates $\Gamma_{-} < \Gamma < \Gamma_{+}$ are likely to
lead to tether formation. For $F_{\rm c} = 20$pN, $\Gamma_{-} \approx
7.2$pN/s and $\Gamma_{+} \approx 435$pN/s. Other parameters used were:
$k_0 = 0.01/$s, $d = 1$nm, $x_{0} = 1\mu$m and $\xi =
1\mu$m/(pNs). (b) Dimensionless parameter regimes in which $P_{\rm
T}\geq 1/2$.  The regions of parameter space below each curve are
associated with $P_{\rm T} > 1/2$, where tether formation is likely.
The smaller the dimensionless bond dissociation rate $\alpha =
k_{0}x_{0}d/(\xi k_{B}T)$, the wider the range of dimensionless
loading rates $\gamma=\xi\Gamma/(k_{0}^{2}x_{0})$ leading to tether
formation. The maximum and minimum pulling rates $\gamma_{-}^{\rm min}
\equiv \gamma_{-}(f_{\rm c}=0)$ and $\gamma_{+}^{\rm max} \equiv
\gamma_{+}(f_{\rm c}=0)$ are indicated for the $\alpha=0.006$ curve,
while the maximal dimensionless delamination force $f_{\rm c}^{\rm
max}$ is shown for $\alpha = 0.004$. (c) The minimum and maximum force
ramps, and the maxima delamination force as functions of $\alpha$. For
$\gamma <\gamma_{-}^{\rm min}$, $\gamma > \gamma_{+}^{\rm max}$, or
$f_{\rm c} > f_{\rm c}^{\rm max}$, tethers {\it always} have less than
50\% chance of forming.}
\label{PTRAMP} 
\end{figure}

When $F_{\rm p}(t) =\Gamma t$, the ligand-receptor survival
probability defined by Eq. \ref{QT} is explicitly

\begin{equation}
Q(t)=\exp\left[-{k_{0}k_{B}T \over \Gamma d}
(e^{\Gamma t d/k_{B}T}-1)\right],
\label{QTRAMP}
\end{equation}
The bond survival probability, evaluated at the time of tether
formation $t_{0}$ (found numerically from Eq. \ref{T0RAMP}),
$Q(t_{0})\equiv P_{\rm T}$, determines the likelihood that a tether is
extracted, and is plotted in Fig. \ref{PTRAMP}(a) as a function of
force loading rate $\Gamma$. Note that $P_{\rm T}$ first increases
with the force loading rate $\Gamma$, before decreasing again at very
high loading rates.  Large critical delamination forces $F_{\rm c}$
increase the probability that ligand-receptor bonds detach before
membrane-cytoskeleton delamination occurs.  Membrane-cytoskeleton
combinations that have weaker delamination forces $F_{\rm c}$ yield a
larger range of force loading rates that lead to tether formation.
Moreover, since the pulling force is specified, $Q(t)$ is independent
of the free tether restoring force $F_{0}$
(cf. Eq.\,\ref{QTRAMP}). The only dependence is on the delamination
force $F_{\rm c}$ which sets the delamination $t_{0}$
(cf. Eq.\,\ref{T0RAMP}) in the expression $P_{T} \equiv Q(t_{0})$.
pnce the tether is formed, the free tether restoring force $F_{0}$ is
irrelevant. The values $\Gamma_{+}$ and $\Gamma_{-}$ are defined by
$P_{\rm T}(\Gamma_{\pm}) = 1/2$ and define the window of loading rates
within which tether formation is likely.

%

%

For quantitative evaluation of $\Gamma_{\pm}$, and their dependences
on the other system parameters ($F_{\rm max}, d,k_{B}T, x_{0}$, and
$\xi$), it is convenient to define dimensionless parameters according
to

\begin{equation}
\begin{array}{lll}
\displaystyle \gamma\equiv {\xi \Gamma \over k_{0}^{2}x_{0}}, &
\displaystyle \alpha \equiv {k_{0} x_{0}d \over \xi k_{B}T}, &
\displaystyle f_{\rm c}= {\xi F_{\rm c}\over
k_{0}x_{0}},\label{NONDIM}
\end{array}
\end{equation}
and find parameter regimes within which $P_{\rm T} \geq 1/2$.  Upon
using Eq. \ref{QTRAMP}, the phase boundaries for tether formation are
determined from the implicit solution to

\begin{equation}
\exp\left[-{1\over
\alpha\gamma}(e^{\alpha\gamma\tau_{0}(\gamma, f_{\rm
c})}-1)\right]={1\over 2},
\label{QHALF}
\end{equation}
where $\tau_{0} = k_{0}t_{0}$ is determined from the solution to
the dimensionless form of Eq. \ref{T0RAMP}:

\begin{equation}
{\gamma\tau_{0} \over f _{\rm c}} - {\gamma\over f_{\rm
c}^{2}}(1-e^{-f_{\rm c}\tau_{0}}) = 1.
\end{equation}
Figure \ref{PTRAMP}(b) shows curves in dimensionless parameter space
($f_{\rm c}$ and $\gamma$) below which $P_{\rm T} > 1/2$.  Asymptotic
analysis of the condition $P_{\rm T} = 1/2$ shows that for
sufficiently large $\gamma$, tether formation will always be
suppressed.  Additionally, as the dimensionless ligand-receptor
dissociation rate $\alpha$ increases, the regime for tether formation
shrinks.

Conditions for tether formation can be further refined by computing
universal parameter curves that define regimes for which $P_{\rm T}$
can never be greater than 1/2.  As a function of the intrinsic
ligand-receptor dissociation rate, there is a band of pulling rates
outside of which $P_{\rm T}$ is always less than one half, even when
tether extraction is barrierless ($f_{\rm c}=0$).
Fig. \ref{PTRAMP}(c) shows $\gamma_{-}^{\rm min}$ and $\gamma_{+}^{\rm
max}$, the minimum and maximum dimensionless ramp rates at which
$P_{\rm T} = 1/2$ when $f_{\rm c} = 0$. These delimiting loading rates
are roots of Eq. \ref{QHALF}. For small $\alpha$, the lower root
$\gamma_{-}^{\rm max} \simeq 2/\ln^{2}2 + O(\alpha)$.

Also plotted in Fig. \ref{PTRAMP}(c) is $f_{\rm c}^{\rm max}(\alpha)$,
the maximum membrane-cytoskeleton delamination force that can give
rise to $P_{\rm T}\geq 1/2$, for any ramp rate. One is unlikely to
pull tethers from membranes that require more force than $f_{\rm
c}^{\rm max}$ to delaminate from the cytoskeleton. Moreover, there is
a critical $\alpha^{\rm max} \approx 0.22444$ above which $P_{\rm
T}=1/2$ cannot be reached, independent of $f_{\rm c}$ or $\gamma$.

\begin{figure}[h!]
\begin{center}
\includegraphics[width=4.0in]{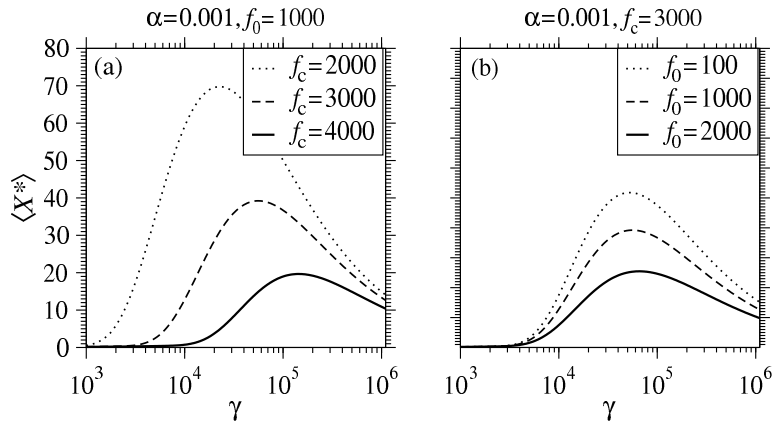}
\end{center} 
\vspace{-2mm}
\caption{(a) Mean tether length $\langle X^{*}\rangle$ as a function
of load rate $\gamma$ for different critical delamination forces
$f_{\rm c}$. (b) $\langle X^{*}\rangle$ at fixed $f_{\rm c}$ and
various $f_{0}$. Even though the cytoskeleton-free membrane restoring
force does not affect tether formation probability $P_{\rm T}$, it
does influence the length of tether possible.}
\label{XSTAR} 
\end{figure}

Finally, we plot in Fig. \ref{XSTAR} the expected dimensionless
maximum tether length $\langle X^{*}\rangle \equiv \langle
x^{*}\rangle/x_{0}$ found from Eq. \ref{meanX}, as a function of the
dimensionless force loading rate $\gamma$.
Not only does the mean tether length decrease with increasing critical
delamination force $f_{\rm c}$, as shown in Fig. \ref{XSTAR}(a), but
the value of $\gamma$ at which $\langle X^{*}\rangle$ is maximized
decreases as $f_{\rm c}$ is raised. Since tether extrusion is a
competition between the rate of climb against the restoring force
$F_{\rm c}x/x_{0}$ and ligand-receptor dissociation, as $f_{\rm c}$ is
increased, higher ramp rates are required to cross the delamination
barrier faster, relative to the bond rupturing.  Although the free
membrane restoring force $f_{0} = \xi F_{0}/(k_{0}x_{0})$ does not
come into play in the tether formation probability, it does influence
the mean length of lipid tether that is extruded before the ligand
detaches from the membrane-bound receptor. Note that in the constant
loading rate protocol, the force is specified and all results are
independent of the pulling device rigidity $K$.



\subsection{Constant Pulling Speed}

Now consider a constant pulling speed protocol. The force felt by the
membrane in this ensemble will depend on the will depend on the
pulling device rigidity $K$.  The bond survival probability is
computed from
 
\begin{equation}
Q(t) = \exp \left[-k_{0} \int_0^t e^{K \ell(t') d/k_{B}T} \dd t' \right],
\label{eq:QofL}
\end{equation}
where $\ell(t)$ is given by Eqs. \ref{eq:LofTsmallT} and
\ref{eq:LofTbigT}.  Initially, while the membrane and cytoskeleton are
attached to each other, and constant speed pulling is applied, the
ligand-receptor bond survival probability $Q(t)$ first decreases
rapidly. After delamination, the force on the ligand-receptor are
fixed, arising only from $F_{0}$ and the viscous drag $\xi^{-1}$.  The
subsequent decay of $Q(t)$ arises from a slower, single exponential.
 


\begin{figure}[h!]
\begin{center}
\includegraphics[width = 4.0in]{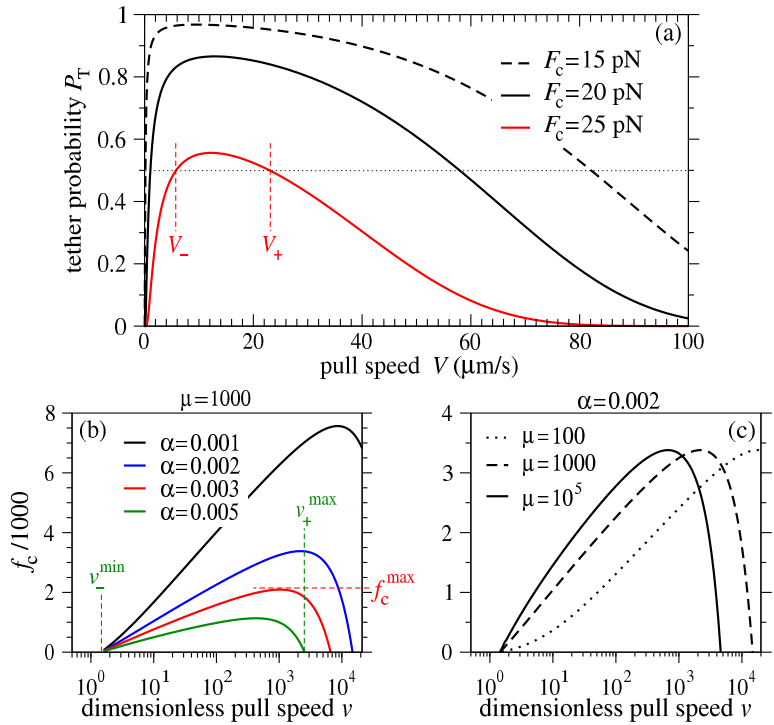}
\end{center} 
\vspace{-2mm}
\caption{(a) The probability $P_{\rm T}$ that tether formation occurs
before the ligand detaches from the receptor plotted as a function of
pulling  speed $V$.  
Analogous to the force ramp protocol, there is a band of pulling
speeds $V_{-} < V < V_{+}$ within which the tether formation
probability is greatest.  Parameters used were: $k_0 = 0.01/$s, $d =
1$nm, $\xi = 1\mu$m/(pN s), $F_{\rm c} = 20$pN, $K = 10$pN/$\mu$m. (b)
Phase diagram in the $f_{\rm c} = \xi F_{\rm c}/(k_{0}x_{0})$ and
$v=V/(k_{0}x_{0})$ parameter space for various
$\alpha=k_{0}x_{0}d/(k_{B}T)$ and fixed pulling device stiffness $\mu
= \xi K/k_{0}= 1000$. (c) Phase diagram for fixed $\alpha = 0.002$ and
various pulling device stiffnesses $\mu$.}
\label{Fig:Pt} 
\end{figure}

Fig. \ref{Fig:Pt}(a) shows the corresponding $P_{\rm T}$ as a function
of pulling speed $V$, and as in the force ramp case, reveals an
optimal pulling speed that maximizes the likelihood of pulling a
tether. In the $V \rightarrow 0$ limit $x(t)\approx 0$ (from
Eqs. \ref{XL} and \ref{eq:LofTbigT}), and we expect $P_{\rm T}
\rightarrow 0$ because when the ligand-receptor bond detaches, the
membrane has not been sufficiently deformed.  In the fast pulling
limit, the detachment rate increases quickly and ligands detach at
very short times $t$ such that tether formation cannot occur. Upon
defining $P_{\rm T}(V_{\pm}) = 1/2$, pulling speeds $V_{-} < V <
V_{+}$ are likely to result in tethers.


To explore how $V_{\pm}$ depends on other system parameters, we employ
the same dimensionless parameters defined in Eqs. \ref{NONDIM}, along
with a dimensionless pulling speed and pulling device stiffness,
\begin{equation}
v \equiv {V \over k_{0}x_{0}}\quad \mbox{and}\quad \mu \equiv {K\xi
\over k_{0}}.
\end{equation}
%

%

Figure \ref{Fig:Pt}(b) shows, for $\mu = 1000$ and various values of
$\alpha$, the boundaries below which tether formation is likely.
Fig. \ref{Fig:Pt}(c) shows the phase boundaries for fixed $\alpha =
0.002$ and various pulling device stiffnesses $\mu$.  Note that softer
pulling devices suppress tether formation at low speeds since the
forces are not immediately felt by the membrane, allowing the ligand
more time to detach. However, softer pulling devices greatly enhance
tether formation at large pull speeds because the accelerated
delamination more than compensates for the drag-mediated acceleration
of ligand-receptor dissociation.  

\begin{figure}[h!]
\begin{center}
\includegraphics[width = 4.3in]{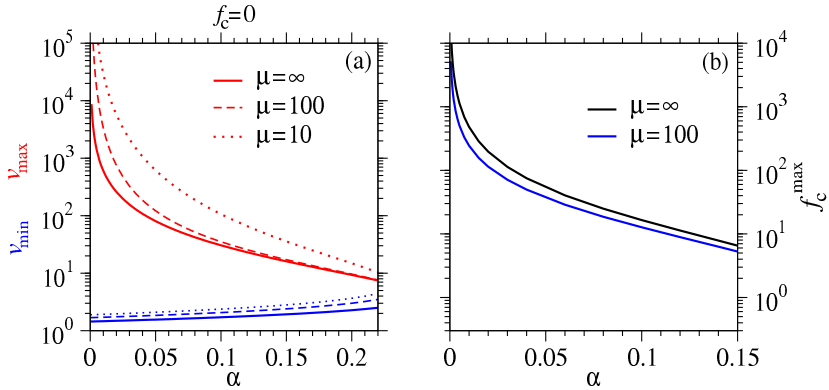}
\end{center} 
\vspace{-2mm}
\caption{(a) Universal curves for the critical values $v_{-}^{\rm
min}$ and $v_{+}^{\rm max}$ as a function of dimensionless unstressed
ligand-receptor dissociation rate $\alpha$. Curves for three
difference values of device stiffness $\mu$ are shown.  For $\alpha
\ll 1$ and $\mu \rightarrow \infty$, the minimum pull speed is
asymptotically $v_{-}^{\rm min} \simeq 1/(\ln 2 -\alpha)$.  In this
limit, the critical $\alpha^{\rm max}$ beyond which $v_{-}^{\rm min}$
and $v_{+}^{\rm max}$ merge (and tether formation is unlikely, even
for $f_{c} = 0$) occurs at $\alpha^{\rm max}\approx 0.24$, slightly
larger than the $\alpha^{\rm max}$ in the constant load rate protocol
(not shown). (b) The maximal delamination force $f_{\rm c}^{\rm max}$
as a function of $\alpha$ for two different stiffnesses $\mu$.}
\label{PHASE_CONSTANT_CRITICAL} 
\end{figure}

Note that analogous to the linear force ramp protocol, for particular
$\alpha$ and $\mu$, there is a maximum and minimum pulling speed
($v_{-}^{\rm min}$ and $v_{+}^{\rm max}$) beyond which tether
formation is unlikely, even when the delamination force vanishes
(Fig. \ref{PHASE_CONSTANT_CRITICAL}(a)). Conversely, there is a
maximum delamination force $f_{\rm c}^{\rm max}$ above which no
pulling speed will result in likely tether formation
(Fig. \ref{PHASE_CONSTANT_CRITICAL}(b)).

\begin{figure}[h!]
\begin{center}
\includegraphics[width=4.0in]{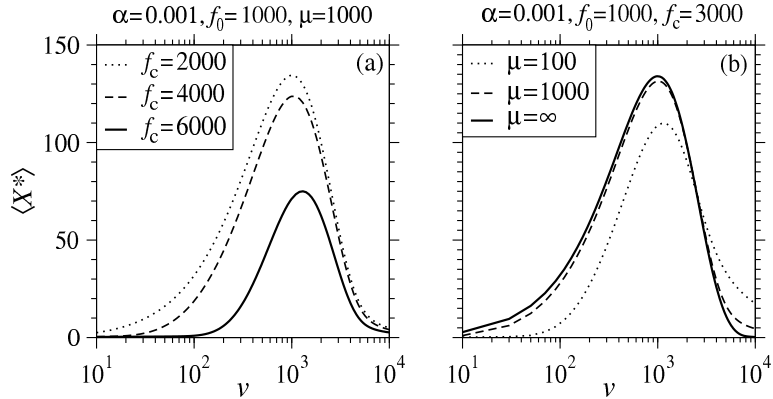}
\end{center} 
\vspace{-2mm}
\caption{(a) The mean system length at ligand detachment, $\langle X^*
\rangle$, is a non-monotonic function of the dimensionless pulling
velocity, $v$, and increases with decreasing $f_{\rm c}$. The
qualitative dependence of $\langle X^{*}\rangle$ on $f_{0}$ is similar
to that shown in Fig. \ref{XSTAR}.  (b) The mean maximum tether length
as a function of $v$ for various dimensionless pulling device
stiffnesses $\mu$.}
\label{XSTAR_CONSTANT} 
\end{figure}

Finally, consider the mean receptor displacement at the moment of
ligand detachment, $\langle X^* \rangle \equiv \langle
x^{*}\rangle/x_{0}$, found from Eq. \ref{meanX} with $X(t) = v\tau
-\ell(t)/x_{0}$, and the appropriate $Q(t)$ and $k_{\rm r}(t)$.  The same
arguments that explain the non-monotonic behavior of $P_{\rm T}$ as a
function of $V$ apply here, and the mean dimensionless receptor
displacement at the moment of ligand detachment, $\langle X^*
\rangle$, is a non-monotonic function of $V$
(Fig. \ref{XSTAR_CONSTANT}). When the pulling velocity is small, the
receptor moves slowly and the term $x(t)$ in Eq. \ref{meanX} will be
small, rendering integral in Eq. \ref{meanX} small.  On the other
hand, when $V$ is sufficiently large, the viscosity $\xi^{-1}$ allows
a large force to be reached, accelerating ligand-receptor bond
rupturing. Thus, $Q(t)$ quickly decreases and the mean receptor
displacement when the ligand detaches, $\langle x^{*}\rangle$, will be
small.

Both constant force load rate and constant pulling speed protocols
give qualitatively similar tether extraction probabilities and maximum
delamination forces $f_{\rm c}^{\rm max}$. This is not surprising
since in the two protocols are physically equivalent in both the
infinitely stiff and infinitely soft pulling device limits, prior to
delamination.  In the large $\mu \propto K$ limit, a constant pulling
speed forces the ligand to have the trajectory $x(t) = Vt$.  Because
we assume a linear force-extension relationship before delamination,
this protocol is equivalent to a high constant force loading rate of
$\Gamma \approx KV$ (or $\gamma \approx \mu v$). In the extremely soft
pulling device limit, the elastic pulling device absorbs most of the
extension and the force at which it acts on the ligand also increases
linearly in time: $\Gamma \approx F_{\rm c}V/x_{0}$ (or $\gamma
\approx vf_{\rm c}$).

However, we do find a qualitative difference in the mean tether length
extracted, due to difference in the post-delamination forces between
the two protocols. As functions of load rate and pulling speed, the
maximum mean tether lengths attainable via linear force ramp are
typically less than half of those achieved through constant pulling
speed, all else being equal. This feature can be understood by
considering how the receptor displacement, $x(t)$ and the ligand
dissociation rate, $k_{\rm r}(t)$ depend on time in each case.  When the
pulling speed is constant, after the tether forms, $x(t)$ increases
linearly in time, and $k_{\rm r}(t)$ is constant.  When we apply a force
ramp to the system, $x(t)$ increases quadratically in time once tether
formation occurs.  This would seem to imply longer tethers under the
force ramp protocol; however, in this case, the dissociation rate
$k_{\rm r}(t)$ also increases exponentially in time.  Thus, ligand
detachment is much faster in the force ramp case, resulting in shorter
observed mean tether lengths.



\section{Summary and Conclusions}

We modeled membrane-cytoskeleton delamination in series with a
ligand-receptor bond and a deformable pulling device and determined
the parameter regimes within which lipid tether extrusion is likely.
Results from our model can be directly used to propose and analyse
experiments in which cell or lipid vesicle membranes are pulled by a
breakable bond. For example, in \cite{SunGrandbois05}, tethers are
pulled from endothelial cells when large force barriers are overcome,
but detachment of the pulling device from the tether is not
considered. Performing such experiments with breakable ligand-receptor
binds would provide the necessary data with which to test our
predictions on the likelihood of tether formation and on the
differences between fixed load rate and pulling speed protocols.

For both linear force ramp and constant pulling speed protocols, we
find a wide window of ramp rates and pulling speeds that likely lead
to tether extraction. However, we also find critical values of a
dimensionless membrane-cytoskeleton delamination force, and a
dimensionless spontaneous ligand-receptor dissociation rate beyond
which tether formation is unlikely, regardless of all other
parameters.  We assumed in all of our analysis that the tether
force-extension curve can be derived from local interactions with a
Helfrich free-energy model.  Finite-size membrane reservoirs and
nonlocal energies such as area-difference elasticity would give rise
to increasing forces as the tether is extended, thereby increasing the
probability of ligand-receptor dissociation, and decreasing expected
tether lengths $\langle X^{*}\rangle$.

Both linear force ramp and constant pulling speed protocols yield
intermediate tether formation regimes, with a specific pulling speed
$v$ and specific linear ramp rate $\gamma$ that maximizes the mean
tether length $\langle X^{*}\rangle$ in the respective protocol.
However, they present different tether dynamics after delamination
leading to different expected tether lengths $\langle
x^{*}\rangle$. Using both protocols, and our results, it may be
possible to characterize membrane-cytoskeleton properties, provided
sufficent information about the ligand-receptor binding energy and
pulling device response are known. In general, such inverse problems
are very ill-posed, but restricting the force-extension relationship
to simple forms as we have done, one may be able to use the onset of
tether formation as a way to estimate force parameters (such as
$F_{0}, F_{\rm c}$ and $x_{0}$).

While we have framed our analysis in terms of AFM experiments in which
the strength of a ligand-receptor bond is probed while the receptor is
in the membrane of a live cell, our basic model is relevant to leukocyte
rolling as well.  In leukocyte rolling, a bond between protein on a
leukocyte microvilli and a protein in the membrane of an endothelial
cell becomes stressed.  Because the microvilli act like Hookean
springs when pulled \cite{ShaoHochmuth98}, and simultaneous extension
of microvilli and tether extraction from the endothelial cells has
been observed \cite{GirdharShao07}, our analysis is directly
applicable to this system, with the cantilever replaced with a
microvilli.

Finally, we note that we have treated the ligand-receptor bond
rupturing as a stochastic Poisson process, while the deformation of
membrane and cytoskeleton was considered deterministic. This
approximation is good as long as $x_{0} \gg d$.  However, if the
experiment is repeated, each region of membrane may have highly
variable attachments to the cytoskeleton. In this case, a distribution
of delamination forces $F_{\rm c}$ should be considered.  Another
source of stochasticity may arise when multiple adhesion points are
being pulled, possibly leading to multiple tethers \cite{MULTIPILI}.
If the entire system is treated as a single, effective tether, the
force-extension of this super-tether will rely on the statistics of
how many individual tethers are still attached during the dynamics.

\vspace{2mm}
 
\ack This work was supported by the NSF through Grant no. DMS-0349195
and by the NIH through Grant no. K25 AI058672.

\vspace{5mm}

\bibliography{pullrefs}

\end{document}